\documentstyle[12pt]{article}

\begin{document}

\def\nn{\nonumber \\}
\def\be{\begin{equation}}
\def\ee{\end{equation}}
\def\ba{\begin{eqnarray}}
\def\ea{\end{eqnarray}}
\def\la{\label}
\def\re{(\ref}

\def\i{{\rm i}}
\let\a=\alpha \let\b=\beta \let\g=\gamma \let\d=\delta
\let\e=\varepsilon \let\z=\zeta \let\h=\eta \let\th=\theta
\let\dh=\vartheta \let\k=\kappa \let\l=\lambda \let\m=\mu
\let\n=\nu \let\x=\xi \let\p=\pi \let\r=\rho \let\s=\sigma
\let\t=\tau \let\o=\omega \let\c=\chi \let\ps=\psi
\let\ph=\varphi \let\Ph=\phi \let\PH=\Phi \let\Ps=\Psi
\let\O=\Omega \let\S=\Sigma \let\P=\Pi \let\Th=\Theta
\let\L=\Lambda \let\G=\Gamma \let\D=\Delta

\def\0{\over } \def\1{\vec } \def\2{{1\over2}} \def\4{{1\over4}}
\def\5{\bar } 
\def\6{\partial }
\def\7#1{{#1}\llap{/}}
\def\8#1{{\textstyle{#1}}} \def\9#1{{\bf {#1}}}

\def\({\left(} \def\){\right)} \def\<{\langle } \def\>{\rangle }
\def\[{\left[} \def\]{\right]} \def\lb{\left\{} \def\rb{\right\}}
\let\lra=\leftrightarrow \let\LRA=\Leftrightarrow
\let\Ra=\Rightarrow \let\ra=\rightarrow
\def\ul{\underline}

\let\ap=\approx \let\eq=\equiv  
        \let\ti=\tilde \let\bl=\biggl \let\br=\biggr
\let\bi=\choose \let\at=\atop \let\mat=\pmatrix
\def\CL{{\cal L}} \def\CD{{\cal D}} \def\rd{{\rm d}} \def\rD{{\rm D}}
\def\CH{{\cal H}} \def\CT{{\cal T}}

\begin{titlepage}
\renewcommand{\thefootnote}{\fnsymbol{footnote}}
\renewcommand{\baselinestretch}{1.3}
\hfill  TUW - 93 - 01 \\
\medskip
\vfill

\begin{center}
{\LARGE {Comment on Gravity and the Poincar\'e Group}}
\medskip
\vfill

\renewcommand{\baselinestretch}{1} {\large {
THOMAS STROBL\footnote{e-mail:
tstrobl@email.tuwien.ac.at} \\ \medskip\medskip
\medskip \medskip
Institut f\"ur Theoretische Physik \\
Technische Universit\"at Wien\\
Wiedner Hauptstr. 8-10, A-1040 Vienna\\
Austria\\} }
\end{center}

\vfill
\begin{center}
submitted to: {\em Phys. Rev. D}
\end{center}
\vfill
\renewcommand{\baselinestretch}{1}                          

\begin{abstract}
Following the approach of Grignani and Nardelli  \cite{Gri}, we show how to
cast the two--dimensional model $\CL \sim curv^2 + torsion^2 + cosm.const$
--- and in fact any theory of gravity ---
into the form of a Poincar\'e gauge theory.
By means of the above example we then clarify the limitations of this
 approach:
The diffeomorphism invariance of the action still leads
 to a nasty constraint algebra.
Moreover, by simple changes
of variables (e.g.\ in a path integral) one can reabsorb all the
modifications of the original theory.
\end{abstract}

\vfill
\hfill Vienna, January 1993  \\
\end{titlepage}

\renewcommand{\baselinestretch}{1}
\setcounter{footnote}{0}

The  similarities of  Cartan's formulation of gravity to the
gauge theories responsible for the remaining interactions has again and again
lead  to
 attempts of reformulating gravity as a gauge theory (cf.\ e.g.\ \cite{Urb}).
The   reformulation of
 pure 2+1 dim  gravity as a Poincar\'e gauge theory with
Chern--Simons action \cite{Wit} and of 1+1 dim Liouville
gravity  as a SO(2,1) ($\L \neq 0$) resp. ISO(1,1) ($\L = 0$)
gauge theory of the BF--type
\cite{Japetc} certainly
spurred such endeavors, all the more since in these cases it was
crucial for the successful quantization.

By introducing the so--called Poincar\'e coordinates $q^a(x^\mu)$
as auxiliary fields, Grignani and Nardelli formulated several
gravitational theories  as Poincar\'e gauge theories \cite{Gri}.
Although their gauge theoretical formulation is
equivalent to the original theories, it,
to our mind,  misses the decisive advantage for quantization
present in the above mentioned works, i.e. the ability to 'eat
up' the diffeomorphism invariance of the respective
gravitational theory by gauge transformations and, correlated to
that, to have to deal with the quite well--known space of flat
connections (cf.\ also
\cite{Bir}). Moreover, as we shall illustrate at the 2 dim model of
NonEinsteinian Gravity given by \cite{Vol}
\be
{\cal L} \, =  e \, (- \frac{\gamma}{4} \, R^2 +
\frac{\beta}{2} \,T^2 - \lambda),
 \la{action}  \ee
it is not only possible to formulate  {\em any} gravitational theory as a
Poincar\'e
gauge theory along the lines of \cite{Gri}, but all of these
formulations are {\em trivially equivalent} to the original ones after an
appropriate shift of variables so that the Poincar\'e coordinates drop
out completely.\footnote{The reader who is further interested in the
classical and quantum mechanical aspects of the integrable model
\re{action}) shall be refered to the literature \cite{Kat}, \cite{All},
\cite{Jap}, \cite{Qua}, as well as references
therein. Here \re{action}) serves only as a nontrivial two--dimensional
example for the present considerations.}

The basic quantities in \re{action}) are
the orthonormal one--forms $e^a$, $e \equiv \det
e_\mu{}^a$, the SO(1,1) connection $\o$, the Ricci
scalar $R = 2 \ast {\rm d} \o$, and
$T^a = \ast \Th^a$ with the torsion two--form $\Th^a \equiv {\rm D} e^a$.
The first order (or Hamiltonian) form of \re{action})
is
\be
\CL_H = -e({\pi_2 \0 2} R + \pi_a T^a - E), \la{LH} \ee
with
\be  E \; \equiv \; \frac{1}{4\gamma} \, (\pi_2)^2
  - \frac{1}{2\b} \, \pi^2 - \lambda,
 \la{E}
\ee
as is most easily seen \cite{Jap} by plugging the field equations for the
momenta $\pi_A \equiv (\pi_a, \pi_2)$ back into \re{LH}).
The first two terms of $\CL_H {\rm d}^2x$ can be
rewritten in a standard manner as $ \pi_A F^A$  in which $F^A$
is the curvature two--form of the appropriate Poincar\'e group ISO(1,1):
$$ F \equiv {\rm d}A + A \wedge A = \Th^a P_a + {\rm d}\o J. $$
This follows by making use of the iso(1,1) Lie Algebra
$[P_a, P_b] =0$, $[P_a,J] = \e_a{}^b P_b$ and setting
\be A = e^aP_a + \o J. \la{A} \ee
The above identifications determine the behavior of the   last term in
\re{LH}) under Poincar\'e transformations, which  then is obvioulsly not
ISO(1,1) invariant. In the spirit of \cite{Gri} we therefore replace it
by \be {1\0 2} \e_{ab} \CD q^a \wedge \CD q^b \, \tilde E \la{hu}\ee
 with $\tilde E$
evolving from \re{E}) by the substitution $\pi_2 \to \ti \pi_2$,
\be \ti \pi_2 \equiv \pi_2 - \e^a{}_b \pi_a q^b. \la{ti} \ee
In \re{hu}) $q^a$ are auxiliary fields transforming under the defining
representation of the Poincar\'e group and $\CD$ is a covariant derivative
ensuring that $\CD q^a$ transformes homogenously, i.e. as a Lorentz vector:
\be \CD q^a \equiv d q^a + \e^a_b \o q^b + e^a =: V^a. \la{V} \ee
The complete ISO(1,1) invariant action density $\ti \CL$ is then
given by ($V \equiv \det V_\m{}^a$):
\be \ti \CL =  \pi_A F^A_{01} + V \ti E. \la{tiL} \ee

That \re{tiL}) is (classically) equivalent to \re{LH}) is already
intuitively clear from the observation that the two Lagrangians
coincide for $q^a=0$, which is an always attainable gauge choice (the
so--called  'physical gauge')
due to $q^a \sim q^a + \rho^a$. Formally it can be verified by
means of second Noether's theorem (cf.\ e.g.\ \cite{Sun})
corresponding to  the above symmetry; one
obtains (for any $S$ with the same symmetry and field content as
$\int d^2x \ti \CL$)
$$ {\d S \0 \d q^a} = \e^b{}_a \o_\m  {\d S \0 \d e_\m{}^b} +
\6_\m {\d S \0 \d e_\m{}^a} +   \e^b{}_a \pi_b {\d S \0 \d \pi_2}  $$
so that the variation with respect to $q^a$ never yields new field equations.
Due to \re{V}), varying  $\ti \CL$ with respect $e^a$, $\o$,
as well as $\pi_A$, and then choosing the physical gauge,
one obviously regains the corresponding variations of   $ \CL_H$.
Thus \re{tiL}) {\em is} a gauge theoretic formulation of 2d NonEinsteinian
Gravity. But does this --- and the other Poincar\'e formulations for
gravitational theories \cite{Gri} except the specific ones mentioned in the
first paragraph --- provide a promising approach for  quantization?

One aspect of the  answer to such a question is provided by a
Dirac--Hamiltonian analysis. \re{tiL}) is already in a first
order form. Instead of applying the procedure suggested in
\cite{Jac1}, however,
 it is in this case more useful to rewrite the term $\e_{ac}
V_1{}^c\ti E \dot q^a$ in $\ti \CL$ as $\, p_a\dot q^a + \l^a(p_a - \e_{ac}
V_1{}^c\ti E )$; denoting the 'rewritten' Lagrangian by $\ti \CL_H$,
which equals $\ti \CL$ when integrating out   $\l^a$ and $p_a$, we obtain
\be \ti \CL_H = \pi_A \dot A_1^A + p_a \dot q^a - \ti \CH \la{tiLH} \ee
with
\ba \ti \CH &=& - A_0^A G_A + \l^a J_a +\6_1 (A_0^A \pi_A)  \la{tiH} \\
    G_a &=&   \partial \, \pi_a - \varepsilon^a{}_b\, A_1^2 \, \pi_a  + p_a \nn
    G_2 &=&   \partial \, \pi_2 +  \varepsilon^a{}_b\, A_1^b \, \pi_a  +
    \e^a{}_b q^b p_a \nn
    J_a &=&  \e_{ab} V_1{}^b \ti E - p_a.  \la{cons}
\ea
 From $\ti \CL_H$
the Hamiltonian structure is obvious: The phase space is spanned by the
canonical coordinates $(A_1^A, q^a; \pi_A, p_a)$ and the (arbitrary) Lagrange
multipliers $A_0^A$,  $\l^a$  enforce the vanishing of the constraints
\re{cons}) [at least when disregarding the surface term in \re{tiH}),
e.g.\ due to periodic boundary conditions in $x^1$].
All of the constraints  are first class:
The $G_A$ are exactly the generators of
ISO(1,1) gauge transformations in the phase space, therefore they satisfy
the corresponding Lie algebra
$$  \{ G_a, G_2 \} = \e_a{}^b G_b \,\d \qquad \{ G_a, G_b\} =0,   $$
$J_a$ behaves as a  Lorentz vector under ISO(1,1) gauge transformations
[cf.\ \re{V})], which leads to
$$  \{ J_a, G_2 \} = \e_a{}^b J_b \,\d \qquad \{ J_a, G_b \} =0,  $$
and a straightforward, somewhat lenghty computation yields
$$  \{ J_a, J_b \} = \e_{ab}  [{1\0 2\g} \ti \pi_2 G_2 -
({1\0 \b} \pi^c + {1\0 2\g}  \pi_2 \e^c{}_d q^d)
G_c  - {1\0 \b}\pi^c J_c ] \,\d.  $$
So in contrast to the gauge theoretical formulations quoted in the first
paragraph of this note here the constraint algebra is not just a
representation  of the Lie algebra of the gauge group. There
appear {\em additional}
constraints responsible for diffeomorphisms, and as usually this
leads to structure {\em functions} of the constraint algebra, one of the
characteristic difficulties of gravity. The reformulation of \re{action})
as \re{tiL}) has by no means simplfied the Hamiltonian structure of the
former (cf.\ \cite{All}),
which is reobtained from the above in the gauge choice $q^a =0$
[this gauge allows to eliminate the $q^a$ and $G_a$ (or $J_a$)
via Dirac brackets,
leaving a phase space spanned by the still conjugates $A_1^A$ and $\pi_A$].

The reason for the appearance of the diffeomorphism constraints
$J_a$ is already obvious from \re{tiL}): On forms a diffeomorphism $x \to
x - f(x)$ acts as a Lie derivative $L_f$, which acting on some connection
one--form A is given by the well--known formula
$$ L_f A = i_fF + D(i_fA). $$
Thus an infinitesimal diffeomorphism can be generated on--shell by a gauge
transformation iff  the field equations enforce
$F=0$.
 This is the
case for $\ti \CL$ only in the limit $\b \to \infty, \g \to
\infty$,\footnote{Let us note on this occasion
that after the identification given in \re{A})
the most complicated 1+1 dim theory of gravity which can be formulated
as a gauge theory with (part of the) field equations being $F=0$
is given by $R=2c_2 =const, \, T^a = c^a = const$.
The corresponding gauge group is defined through
$[P_a, P_b] =\e_{ab}(c_2 J + c^dP_d)$, $[P_a,J] = \e_a{}^b P_b$, reproducing
\cite{Japetc} for the case $c^a =0$. Also when allowing for more then three
generators (cf.\ \cite{Jac}), it will never be possible to formulate
\re{action}) as a BF--theory.}
which for $\l = 0$
reproduces just the Liouville gravity with vanishing cosmological
constant, mentioned in the first paragraph
(for $\l \neq 0$ cf.\ below).

In our context even more striking, however,
is the following observation.
Using the definition for $\ti \pi_2$ and $V^a$ one easily
verifies
\be \pi_A F^A \equiv  \pi_a (d e^b + \e^a{}_b \o \wedge e^b) + \pi_2 d \o
   =   \pi_a (d V^b + \e^a{}_b \o \wedge V^b) + \ti  \pi_2 d \o
   \la{equ} \ee
so that we find
\be \ti \CL (e^a,\o,\pi_a,\pi_2,q^a) =
\CL_H (V^a,\o,\pi_a,\ti \pi_2) \sim \CL(V^a,\o).  \la{tri} \ee
That is to say
the appearance of $e^a$ and $q^a$ within $\ti \CL$ can be reabsorbed
into the 'true vielbein' $V^a$ and a redefinition of the field $\pi_2$;
integrating out $\pi_A$, moreover,
one ends up with \re{action}) in
which $e^a$ has been replaced by $V^a$. At the classical
level  this redefinition \re{ti}), \re{V})
of coordinates can be performed either before or
after a minimalization of the action. But also
within a path integral the
corresponding functional determinant yields just
 one.  Therefore the Poincar\'e gauge theoretic
formulation of \re{action}), given in \re{tiL}),
reduces to a mere renaming of $e^a$ by $V^a$.

That this is not a special feature of the present model \re{action})
shall be illustrated by means of the action $S_S$ of a scalar field
$\ph$ coupled to 4 dim Poincar\'e gravity which was given in the first
ref.\ \cite{Gri}:
$$  S_S = - {1\0 3} \int_{M_4}  \e_{abcd} \CD q^a \wedge \CD q^b
  \wedge \CD q^c \wedge (\ph \CD \ph^d - \ph^d d \ph + \ti \ph^e
  \ti \ph_e \CD q^d)   $$
 with (the Lorentz vectors)
\ba \CD \ph^a &=& d \ph^a + \o^a{}_b\, \ph^b + m^2 e^a \ph \,, \nn
 \ti \ph^a &=& \ph^a - m^2 q^a \ph.  \la{umb} \ea
Although not obvious at first sight,
 shifting  the {\em auxiliary} field $\ph^a$ according
to \re{umb}),
also here $e^a$ and $q^a$ can be recombined to the combination
$V^a$ given in \re{V}).  One obtains
$$ S_S = - {1\0 3} \int_{M_4}  \e_{abcd} V^a \wedge V^b
  \wedge V^c \wedge [\ph D \ti \ph^d -  \ti \ph^d d \ph + (\ti \ph^e
  \ti \ph_e + m^2 \ph^2)V^d  ]  $$
with the Lorentz covariant derivative
$$ D \ti \ph^a = d \ti \ph^a + \o^a{}_b \ti \ph^b.  $$

As a byproduct of these considerations
let us note the incorrectness of the statement in the
appendix of the second ref.\ \cite{Gri},
namely  that always $V \neq 0$ in the case of
2d black hole gravity in its  Poincar\'e formulation. The latter is given
by \re{tiL}) in the limit $\b \to \infty, \g \to \infty$. According to
 \re{tri})  this Lagrangian is equal to \re{LH})
 in the same limit [which
 does not affect \re{ti}), \re{V})] when exchanging
$e^a$ by $V^a$. Since this Lagrangian in turn
can be understood as a Poincar\'e gauge theory \cite{Ver} with connection
\re{A}), $V^a \lra e^a$, --- the cosmological constant term yields only a
surface term under ISO(1,1) transformations ---
any axial gauge, obtainable at least locally, leads to $V=0$.

Clearly one can vice versa obtain a Poincar\'e gauge theory from any
theory of gravity by the replacement $e^a := \CD q^a$, or, even simpler,
one can regard any theory of gravity as being already a Lorentz gauge theory
when allowing for {\em one}--forms $e^a$ in the fundamental
representation. However, this does not seem to provide any
advantage  in
the quest for quantization of gravity.

The author wants to  thank W. Kummer and D. J. Schwarz for reading the
manuscript.

\end{document}